\documentclass[12pt, leqno]{article}
\newtheorem{prop}{Proposition}
\newtheorem{hyp}{Hypothesis}
\newtheorem{lem}{Lemma}
\newtheorem{rem}{Remark}
\newtheorem{cor}{Corollary}
\newtheorem{thm}{Theorem}
\newcommand{\be}{\begin{equation}}
\newcommand{\ee}{\end{equation}}
\newcommand{\ba}{\begin{array}}
\newcommand{\ea}{\end{array}}

\newcommand{\curl}{{\rm \; curl\;}}

\newcommand{\bea}{\begin{eqnarray*}}
\newcommand{\eea}{\end{eqnarray*}}
\newcommand{\bean}{\begin{eqnarray}}
\newcommand{\eean}{\end{eqnarray}}
\newcommand{\proof}{\vspace{1ex}\noindent{\em Proof}. \ }
\def\ds{\displaystyle}
\def\nm{\noalign{\medskip}}
\newcommand{\C}{\mbox{\bf C}}
\newcommand{\R}{\mbox{\bf R}}
\def\Box{\leavevmode\vbox{\hrule
     \hbox{\vrule\kern5pt\vbox{\kern5pt}%
           \vrule}\hrule}}
\newcommand{\square}{\hfill$\Box$}
\begin{document}

\title{Unique determination of electromagnetic parameters from partial boundary measurements}

\author{Christian Daveau \thanks{
 Department of Mathematics, CNRS (UMR 8088),
 University of Cergy-Pontoise,
 2 avenue Adolphe Chauvin,
 95302 Cergy-Pontoise Cedex, France. (Email: christian.daveau@u-cergy.fr)}
%\and Maroua Gozzi
%\thanks{ Departement of Mathematics, University of Carthage, Bizerte, Tunisia
%(Email:gozzi.maroua@live.fr)}
\and Abdessatar Khelifi \thanks{
D\'epartement de Math\'ematiques, Universit\'e de Carthage, Tunisie. (Email: abdessatar.khelifi@fsb.rnu.tn)}\and Houssem Lihiou \thanks{ Department of Mathematics, Faculty of Sciences, 7021 Zarzouna, Bizerte, Tunisia.
(Email: houssemlihiou@gmail.com)} }

\maketitle \abstract{We consider an inverse boundary value problem for the Maxwell's equations with a given data assumed to be known only in accessible part $\Gamma$ of the boundary. We aim to prove an uniqueness result using the Dirichlet to Neumann map with measurements limited to an open part of the boundary and we seek to reconstruct
 the complex refractive index $\emph{\textbf{n}}$ in the interior of a bounded domain Further, using the impedance map restricted to $\Gamma$, we may identify locations of small volume fraction
perturbations of the refractive index.}\\


\noindent {\footnotesize {\bf Mathematics Subject
Classifications:  (MSC2010)} 35Q61, 78M35, 35R30

%%%%%%%%%%%%%%%%%%%%%%%%%%%%%%%%%%%%%%%%

\section{Introduction}
In this paper, we consider an inverse boundary value problems for the time-harmonic Maxwell's equations in a bounded domain, that is, to reconstruct specific electromagnetic parameter: complex refractive index $\emph{\textbf{n}}(x)$, as function of the spatial variable, from a specified set of partial electromagnetic field measurements taken on the boundary.\\
For a closely related problem to the one considered here, we refer the readers to the original work of Colton and P$\ddot{a}$iv$\ddot{a}$rinta in \cite{CP}. The authors showed that the
refractive index $\emph{\textbf{n}}(x)$ (corresponding to e.g., known constant $\mu$ but unknown $\varepsilon(x)$ and
$\sigma(x)$) can be uniquely determined by the far-field patters of scattered electric fields
satisfying time-harmonic Maxwell's equations. Their approach is based on the ideas, developed by Sylvester and Uhlmann in \cite{SU}, of constructing CGO (Complex Geometric Optic) type of solutions. In this context, the unique recovery of electromagnetic parameters from the scattering amplitude was
first proven in \cite{CP} under the assumption that the magnetic permeability $\mu$ is a constant. But, the unique recovery of general $C^2$-electromagnetic parameters
 $\mu$ and $\gamma$ from full boundary data was later proved in \cite{OPS1}, and simplified in \cite{OS} by introducing the so-called
generalized Sommerfeld potentials. Concerning boundary determination results, we may refer to \cite{COS,JM,KV,McDowall1,McDowall2,Khelifi1} and \cite{Uhlmann2}. For a slightly more general approach and more background information, see also \cite{OPS2}.\\
Inverse problems with partial data for scalar elliptic equations have attracted considerable
attention recently. In \cite{BU,KSU}, the authors use Carleman estimates in their approaches. In \cite{Ammari-Uhlmann}, the authors use the local Dirichlet-to-Newmann map to recover the unknown coefficient by measuring on part of the boundary, but in \cite{isakov} the author uses reflection arguments.\\
%%%%%%%%%%%
However, in electromagnetic problems concerned with the partial data problem, namely, to determine the parameters from the
impedance map only made on part of the boundary, there are not as many results as
in the scalar case. It is shown in \cite{COS} that if the measurements $\Lambda(f)$ is taken only on a
nonempty open subset $\Gamma$ for $f = \nu\times E|_{\partial\Omega}$ supported in $\Gamma$, where the inaccessible
part $\overline{\partial\Omega\backslash\Gamma}$ is part of a plane or a sphere, the electromagnetic parameters can still be
uniquely determined. Combined with the augmenting argument in \cite{OS}, the proof in \cite{COS}
generalized the reflection technique used in \cite{isakov}. As for another well-known method in dealing with partial
data problems based on the Carleman estimates \cite{BU,KSU}, there are however significant
difficulties in generalizing the method to the full system of Maxwell's equations, e.g.,
the CGO solutions constructed using Carleman estimates.\\
The novelty of this paper lies in the use of partial electromagnetic field measurements taken on accessible part $\Gamma$ to recover a complex refractive index. These partial measurements are traduced by an exhibition of appropriate partial boundary measurements in the form of a restricted boundary mapping $Z$, specifically the mapping from the
tangential components of the electric field $\textbf{E}$ to the tangential components of $\mbox{curl }\textbf{E}$ on the nonempty part $\Gamma$. In this article, we consider the mapping $Z$ which is closer to a natural generalization of the resistive map considered in
impedance imaging applications. Our ideas and our methods differ from the approaches developed by Brown et al. in \cite{Brown} or by Caro et al. in \cite{COS,caro1}. Other inverse problems in electromagnetism in settings different to the ones in this paper have been considered in \cite{Ammari-Kang,AVV,Salo1,Kurylev1,Kurylev2,Liu1,Masahiro1,Masahiro2,Romanov,darbas,Khelifi1}.\\

The outline of the paper is as follows. In the next section we recall some useful notation and function space, and we formulate the underlined problem. In Section $3$, We eliminate the magnetic field and we reduce previous Maxwell's equations to a system of equations for electric field ${\bf E}$. The global uniqueness result is provided. In Section $4$, we derive a formula for calculating the unknown refractive index $\emph{\textbf{n}}$ from the local impedance map $Z_{\emph{\textbf{n}}}$, and by the Fourier integral theorem. We conclude our paper in Section $5$ by applying the reconstruction procedure described before for identifying the locations of small volume fraction perturbations of the refractive index.
%%%%%%%%%%%%%%%%%%%%%%%%%%%%%%%%%%%%%%%%%%%%%%%%%%%%%%%%%%%%%%%%%%%%%%%%%%%%%%%%%%%%%%%%%%%%%%%%%%%%%%%%%%%%%%%%%

%%%%%%%%%%%%%%%%%%%%%%%%%%%%%%%%%%%%%%%%%%%%%%%%%%%%%%%%%%%%%%%%%%%%%%%%%%%%%%%%%%%%%%%%%%%%%%%%%%%%%%%%%%%%%%%%%%%%%%%%%%%%%%%%%

\section{Problem formulation}
%\subsection{Notation}
In the present paper, the following notation is used. If $\mathcal{F}$ is a function space,
$\mathcal{F}^p$, $p\in \textbf{N}$, denotes the space of vector-valued functions with each of the $p$ components in $\mathcal{F}$. The usual $\textbf{L}^2$-based Sobolev spaces are denoted by $H^s$, $s\in \R$. On the boundary of $\partial\Omega$, the Sobolev spaces of tangential fields are defined as
$$TH^s(\partial\Omega)=\{\textbf{f}\in(H^s(\partial\Omega))^3;\quad \textbf{f}\cdot\nu=0\}.
$$
Here, $\nu=\nu(x)$ is the exterior unit normal vector of the boundary at $x\in\partial\Omega$. We also
need spaces of tangential fields having extra regularity. Let "div" denote the surface
divergence on $\partial\Omega$ (see, e.g., \cite{kress} or \cite{Nedelec} for the definition). We define
$$TH_{\mbox{div}}^s(\partial\Omega)=\{\textbf{f}\in TH^s(\partial\Omega);\quad \mbox{div }\textbf{f}\in H^s(\partial\Omega)\}.
$$
Finally, we remind that $TH_{\mbox{div}}^s(\partial\Omega)$-spaces arise naturally through the tangential trace
mapping acting on functions in the spaces of the type
$$H_{\mbox{div}}^s(\Omega)=\{\textbf{f}\in (H^s(\Omega))^3;\quad \mbox{div }(\nu\times\textbf{f}|_{\partial\Omega})\in H^{s-1/2}(\partial\Omega)\}.
$$
Here $\times$ stands for the vector product. The div-spaces are discussed to some extent in the references \cite{BCSheen}, \cite{kress} and \cite{SIC}.\\
Throughout this paper, we use "$\cdot$" to denote the standard scalar product in $\R^3$.\\

%%%%%%%%%%%%%%%%%%%%%%%%%%%%%%%%%%%%%
%\subsection{Statement of the problem}
Let $\Omega \subset \R^3$ be a nonempty, open, and bounded set having a $C^2$-smooth boundary $\partial\Omega$. The unit
normal vector to $\partial\Omega$, which is directed into the exterior of $\Omega$, is denoted by $\nu$. Moreover, we
assume that the exterior domain $\Omega_e :=\R^3\backslash \overline{\Omega}$ is connected. Let
$\Gamma$ be a smooth open subset of the boundary $\partial \Omega$
and $\Gamma_c$ denotes $\partial \Omega \setminus
\overline{\Gamma}$.\\

Consider first the boundary value problem of finding the electromagnetic fields $\textbf{E}$ and $\textbf{H}$ in a non-magnetic medium of bounded support $\Omega$:
\begin{equation}\label{equation1}\mbox{curl }\textbf{E}-i\omega\mu \textbf{H}=0,\quad \mbox{curl }\textbf{H}+i\omega\varepsilon_0\emph{\textbf{n}}(x) \textbf{E}=0,\quad \mbox{in }\Omega,
\end{equation}
with the electric boundary condition\begin{equation}\label{equation2}\nu\times \textbf{E}|_{\partial \Omega} = \textbf{f}\in TH_{\mbox{div}}^{1/2}(\partial \Omega ).
\end{equation}
Physically, $(\textbf{E}, \textbf{H})$ is the time-harmonic electromagnetic field, $\omega>0$ is its (fixed) frequency, \begin{equation}\label{index}\emph{\textbf{n}}(x)=\frac{\varepsilon(x)}{\varepsilon_0}+i\frac{\sigma(x)}{\omega\varepsilon_0}\end{equation}
 is the refractive index of the medium, $\varepsilon$ denotes the electric permittivity of the conductor $\Omega$, $\mu$ denotes the
magnetic permeability of $\Omega$, and $\sigma$ denotes its conductivity.\\
%%%%%%%%%
Moreover, we recall that $\varepsilon_0>0$ and $\mu_0>0$ are respectively the permittivity and the permeability in the vacuum.\\
We assume the following conditions on the material parameters.
\begin{hyp}
The permittivity $\varepsilon$, permeability $\mu$, and conductivity $\sigma$ are $C^2$
functions verifying the following properties.
\begin{itemize}
\item[-] The magnetic permeability $\mu(=\mu_0)$ is constant in this non-magnetic medium.
  \item[-] For some positive constants $\varepsilon^{-}$, $\varepsilon^{+}$ and $\sigma^{+}$,
$$0<\varepsilon^{-}\leq \varepsilon(x)\leq \varepsilon^{+},\quad 0\leq \sigma(x)\leq\sigma^{+};\quad \mbox{for }x\in\overline{\Omega}.
$$
  \item [-]  The function $\sigma$ and the difference $\varepsilon-\varepsilon_0$ are in $C_{0}^{2}(\Omega)$.
\end{itemize}
\end{hyp}
Now, under above properties we can claim that the refractive index $\emph{\textbf{n}}$ is a complex function satisfying $\emph{\textbf{n}}\in C^{2}(\bar{\Omega})$ for some $0 < \alpha < 1$, $Re( \emph{\textbf{n}} )> 0,$ $Im (\emph{\textbf{n}}) \geq0$. Moreover, if we denote by $\emph{\textbf{n}}_0$ the refractive index in $\Omega_e :=\R^3\backslash \overline{\Omega}$, then we have $\emph{\textbf{n}}-\emph{\textbf{n}}_0\in C_0^2(\Omega)$.\\

It is known \cite{OS} that the above boundary value problem (\ref{equation1})-(\ref{equation2}) has a unique solution $(E, H)\in H_{\mbox{div}}^{1}(\Omega )\times H_{\mbox{div}}^{1}(\Omega )$  except for a discrete set of electric resonance frequencies ${\omega_n}$ when $\sigma\equiv0$. Assuming that $\omega$ is not a resonance frequency. Then according to \cite{OS,OPS1,SIC,Uhlmann1}, and by considering electric field $\textbf{E}$ instead of the magnetic $\textbf{H}$ which considered in the previous references (see for example \cite{OS}), we can state that the following map\begin{equation}\label{equation3}Z: TH_{\mbox{div}}^{1/2}(\partial \Omega )\to TH_{\mbox{div}}^{1/2}(\partial \Omega ),\quad \textbf{f}\mapsto\nu\times\textbf{H}|_{\partial \Omega}
\end{equation}
called \emph{the impedance map} is well defined.\\
Recall that,
\begin{equation}\label{relation-E-H}{\bf H}= \ds\frac{1}{i
\omega \mu_{0}}
 \mbox{curl }{ \bf E},\end{equation}
and scaling in (\ref{equation3}) by the complex constant $i\omega \mu_{0}$. Then, the following map \begin{equation}\label{equation4}Z: TH_{\mbox{div}}^{1/2}(\partial \Omega )\to TH_{\mbox{div}}^{1/2}(\partial \Omega ),\quad \textbf{f}\mapsto\nu\times\mbox{curl }\textbf{E}|_{\partial \Omega},
\end{equation}
which denoted also by $Z$ and still called \emph{the impedance map}, is well defined.\\
%remarque: on peut faire ca ou impedance local sous forme proposition.... sa preuve repose sur (8) et result sur impedance en references...
It is clear, that if $\Omega$ is magnetic medium ($\mu=\mu(x)$), the mapping defined in (\ref{equation4}) can not be comparable to those in the literature. This requires more attention and analysis, but we remove this consideration in the present paper.

%%%%%%%%%%%%%%%%%%%%%%%%%%%%%%%%%%
\section{Global uniqueness}
%%%%%%%%%%%%%%%%%%%%%%%%%%%%%%%%%%%%%%%%
We, now, eliminate the magnetic field from the equations (\ref{equation1})-(\ref{equation2}) by
dividing the first equation in~(\ref{equation1}) by $i\omega\mu$ and taking the curl to obtain the following system of equations for electric field ${\bf E}$:
\begin{eqnarray}\label{eq-maxwell1}
  \mbox{curl }\mbox{curl }\textbf{E} - k^2\emph{\textbf{n}}\textbf{E}=0 \quad  \mbox{in }\Omega&& \\
  \nu\times \textbf{E} = \textbf{f}\quad  \mbox{on }\partial\Omega,&& \\\nonumber
\end{eqnarray}
where $k = \omega\sqrt{\mu_0\varepsilon_0}$ is the wave number corresponding
to the background, $\emph{\textbf{n}}$ the refractive index defined by (\ref{index}), and $\textbf{f}\in TH_{\mbox{div}}^{1/2}(\partial \Omega )$ is a given data on $\partial\Omega$.\\

Introduce the trace space \be\label{tildeH}\widetilde{TH}_{\mbox{div}}^{1/2}(\Gamma): =
\Bigr\{ \textbf{f} \in TH_{\mbox{div}}^{1/2}(\partial\Omega), \textbf{f} \equiv 0
\mbox{ on } \Gamma_c \Bigr\}.\ee
 Here and in the sequel we identify
$\textbf{f}$ defined only on $\Gamma$ with its extension by $0$ to all
$\partial \Omega$ ($supp(\textbf{f})\subset\Gamma$).
%Our main result in this section is\\
\begin{rem}\label{rem1}Let $\it{\textbf{n}}\in C^{2}(\bar{\Omega})$, and suppose that $k^2$ is not an eigenfrequency of the following problem:
$$\left\{
\begin{array}{ll}
(\mbox{curl }\mbox{curl } - k^2\it{\textbf{n}})\textbf{E}=0, & \hbox{in }\Omega \\
\nu\times \textbf{E}|_{\partial\Omega} = \textbf{f}\in\widetilde{TH}_{\mbox{div}}^{1/2}(\Gamma), &
\end{array}
\right.$$
where $\widetilde{TH}_{\mbox{div}}^{1/2}(\Gamma)$ was defined by (\ref{tildeH}). Then, $Z_{\it{\textbf{n}}}(\textbf{f})=\nu\times\mbox{curl }\textbf{E}|_{\Gamma}$ is the {\em local impedance map} in this
case.\\
\end{rem}
Based on definition (\ref{equation4}) and on Remark \ref{rem1}, the inverse boundary value problem is to recover $\emph{\textbf{n}}$ from the partial
boundary measurements encoded as the well-defined local impedance map:
\begin{equation}\label{local-impedance}Z_{\emph{\textbf{n}}}: \widetilde{TH}_{\mbox{div}}^{1/2}(\Gamma)\to TH_{\mbox{div}}^{1/2}(\partial \Omega ),\quad \textbf{f}\mapsto\nu\times\mbox{curl }\textbf{E}|_{\Gamma}.\\
\end{equation}

We will prove the following main result, showing that partially boundary measurements
for the Maxwell's equations uniquely determine the material parameters in
a bounded domain.
\begin{thm} \label{th1}Let $\it{\textbf{n}}_0$ be the refractive index in $\R^3\backslash \overline{\Omega}$ such that $\it{\textbf{n}}_i-\it{\textbf{n}}_0\in C_0^2(\Omega)$, for $i=1, 2.$
Assume $\it{\textbf{n}}_1 =\it{\textbf{n}}_2$ almost everywhere in a neighborhood of the
boundary $\partial \Omega$. Suppose that $k^2$ is not a resonant frequency for (\ref{eq-maxwell1})-(8) associated to $\it{\textbf{n}}_i,$ $i=1, 2$. If the local impedance maps
coincide,$$Z_{\it{\textbf{n}}_1}(\textbf{f}) = Z_{\it{\textbf{n}}_2}(\textbf{f})\quad \mbox{for all }\textbf{f}\in \widetilde{TH}_{\mbox{div}}^{1/2}(\Gamma),
$$
then there exists $\kappa=\kappa(\Omega)>0$ such that $$\it{\textbf{n}}_1=\it{\textbf{n}}_2$$
almost everywhere in $\Omega$ whenever
 $$\|\it{\textbf{n}}_j-\it{\textbf{n}}_0\|_{W^{2,\infty}(\Omega)}<\kappa.
 $$
\end{thm}
Before proceeding with the proof of Theorem \ref{th1}, we remind some well-know and original results.\\
%%%%%%%%%%%%%%%%%%%%%%%%%%%%%%%%%%%
By referring to the works of Colton and P$\ddot{a}$iv$\ddot{a}$rinta \cite{CP}, Sun and Uhlmann \cite{Uhlmann1} and to the Calder\'on problem of electrostatics (see e.g., \cite{C,N} and \cite{SU}), we consider the inverse problem of determining the key electromagnetic parameters from the
boundary measurement. Particularly, in this paper we assume that we can measure the values of $Z_\emph{\textbf{n}}(\textbf{f})$ only on a nonempty open subset $\Gamma\subset\partial\Omega$, and only for tangential boundary fields $\textbf{f}$ supported in $\Gamma$.\\

%%%%%%%%%%%%%%%%%%%%%%%%%%%%%%%%%%%%%%%%%%%%%%%
The common outline of the proof of Theorem \ref{th1} follows approximately the same lines as the
proof of the global uniqueness theorem for the inverse conductivity problem
given in \cite{SU}, for an inverse boundary value problem for Maxwell's
equations in \cite{Uhlmann1}, for the uniqueness of a solution to
an inverse scattering problem for electromagnetic waves in \cite{CP} or for proof of the global uniqueness theorem for the Schr$\ddot{o}$dinger equation in \cite{Ammari-Uhlmann}.
Explicitly, one first proves an identity involving products of
solutions of the equation under consideration as will be done in Lemma \ref{lem1}. Next, one proves a density result as in Lemma \ref{lem2}. Then one constructs vector CGO type solutions
 for the underlined problem (\ref{eq-maxwell1})-(8) to obtain information, via this identity,
of the Fourier transform of the unknown function. There are two main difficulties
in carrying out this approach for the problem under consideration
here. First, we cannot reduce Maxwell's equations to a Schr$\ddot{o}$dinger equation to proceed exactly as in \cite{Ammari-Uhlmann} (for example).
The best we can do is to reduce Maxwell's equations to a system whose
principal part is the Laplacian times the identity operator as done in \cite{CP,Uhlmann1,OS,OPS1,OPS2}. We can then construct
CGO solutions under appropriate smallness assumptions. Also, in our case we have to construct global solutions in
order to guarantee that the solutions constructed satisfy the condition that the
electric and magnetic field be divergence-free. In order to determine the
unknown $\emph{\textbf{n}}$, one has to study the asymptotic expansion
of these solutions in a free parameter. The second difficulty
is that such CGO solutions for Maxwell's equations do not have the property
that $R_\xi$ decays like $O(1/|\xi|)$ (see e.g., \cite{CP,OS,OPS1,OPS2}), which was a key ingredient in the proof of the uniqueness
in the scalar case. But, this is tackled in \cite{CP} by constructing appropriate $R_\xi$ that decays to zero in certain distinguished directions as $|\xi|$ tends
to infinity. By carefully choice of several directions for $\xi$, as will be defined in relation (\ref{eqv2-1}), such special set of solutions are enough to determine the
refractive index.\\

To prove Theorem \ref{th1}, we begin by the following lemma. This result generalizes the Alessandrini's identity \cite{A} for the
conductivity equation to Maxwell system.\\
%Let $\Omega_0:=supp(\emphn{\textbf{n}}(x)-\emph{\textbf{n}}_0)$ such that $\Omega_0\subset \Omega$.
\begin{lem} \label{lem1}
Let $\mathcal{O} \subset \subset \Omega$ containing $supp(\it{\textbf{n}}_j - \it{\textbf{n}}_0)$ (for $j=1,2$), such that $\mathcal{O}$ is a bounded domain with $C^2$ boundary. Let
$\textbf{E}_i \in \textbf{H}_{\mbox{div}}^1(\Omega)$ satisfying:
\be
\label{eq0}\begin{array}{l} (\frac{1}{k^2}\mbox{curl }\mbox{curl }- \it{\textbf{n}}_i)\textbf{E}_i= 0  \quad  \mbox{in }\Omega \\
 \nu\times \textbf{E}_i|_{\Gamma_c} = 0, \quad i=1, 2\\
 \nu\times \textbf{E}_i|_{\partial\mathcal{O}} \neq 0, \quad i=1, 2.
\end{array}
\ee Assume $\it{\textbf{n}}_1 =\it{\textbf{n}}_2$ almost everywhere in a neighborhood of the
boundary $\partial \Omega$ and $Z_{\it{\textbf{n}}_1} = Z_{\it{\textbf{n}}_2}$. Then, $supp(\it{\textbf{n}}_1-\it{\textbf{n}}_2)\subset \mathcal{O}$ and
\be
\label{i2} \ds \int_{\mathcal{O}}  \textbf{E}_1\cdot (\it{\textbf{n}}_1 -\it{\textbf{n}}_2)\textbf{E}_2 \; dx = 0.
\ee
\end{lem}
\proof Firstly, to get $supp(\emph{\textbf{n}}_1-\emph{\textbf{n}}_2)\subset \mathcal{O}$, one may expand $$(\emph{\textbf{n}}_1-\emph{\textbf{n}}_2)(x)=(\emph{\textbf{n}}_1-\emph{\textbf{n}}_0)(x)-(\emph{\textbf{n}}_2-\emph{\textbf{n}}_0)(x),$$and the result follows immediately by recalling that $supp(\emph{\textbf{n}}_j - \emph{\textbf{n}}_0)\subset \mathcal{O}$ for $j=1,2$.\\
 Now, let $\textbf{E}_i$, $i=1,2$ be solutions of (\ref{eq0}). Then, by using Green's theorem we have that

$$\ds \int_{\Omega} \mbox{curl }\textbf{V}\cdot\textbf{E}_i \; dx =\int_{\Omega} \textbf{V}\cdot\mbox{curl }\textbf{E}_i \; dx
+\int_{\partial\Omega} (\nu\times\textbf{V})\cdot\textbf{E}_i \; ds(x),
$$where $ds(x)$ denotes
surface measure.\\

If $\textbf{V}|_{\Gamma_c}\equiv 0$, the above relation becomes
 $$\ds \int_{\Omega} \mbox{curl }\textbf{V}\cdot\textbf{E}_i \; dx =\int_{\Omega} \textbf{V}\cdot\mbox{curl }\textbf{E}_i \; dx
+\int_{\Gamma} (\nu\times\textbf{V})\cdot\textbf{E}_i \; dx.
$$
Therefore, by replacing $\textbf{V}:=\mbox{curl }\textbf{E}_i$ and using Green's theorem for $i=1$ and $i=2$ respectively, relation (\ref{eq0}) gives
\be
\label{eq1-0} \ds \int_{\Omega} \textbf{E}_1\cdot (\emph{\textbf{n}}_1 -\emph{\textbf{n}}_2)\textbf{E}_2  \; dx =\frac{1}{k^2}
\int_\Gamma [(\nu\times\mbox{curl }\textbf{E}_1) \cdot\textbf{E}_2 \ee
$$-\textbf{E}_1\cdot(\nu\times\mbox{curl }\textbf{E}_2)  ]\; ds , $$
Recall that $supp(\emph{\textbf{n}}_1 -\emph{\textbf{n}}_2)\subset\mathcal{O}\subset\Omega$, then from (\ref{eq1-0}) we immediately get
%%%%%%%%%
\be
\label{eq1} \ds \int_{\mathcal{O}} \textbf{E}_1\cdot (\emph{\textbf{n}}_1 -\emph{\textbf{n}}_2)\textbf{E}_2  \; dx =\frac{1}{k^2}
\int_\Gamma [(\nu\times\mbox{curl }\textbf{E}_1) \cdot\textbf{E}_2 \ee
$$-\textbf{E}_1\cdot(\nu\times\mbox{curl }\textbf{E}_2)  ]\; ds . $$

On the other hand, let $\textbf{V}_1 \in \textbf{H}^1(\Omega)$ be solution of $(\frac{1}{k^2}\mbox{curl }\mbox{curl }- \emph{\textbf{n}}_1)\textbf{V}_1$
in $\Omega$ such that $\nu\times\textbf{V}_1 |_{\Gamma_c} =0$ and $\nu\times\textbf{V}_1 |_\Gamma =
\nu\times\textbf{E}_2 |_\Gamma$.\\ From $\Lambda_{\emph{\textbf{n}}_1} = \Lambda_{\emph{\textbf{n}}_2}$ it follows that \be
\label{a1} \ds\nu\times \textbf{V}_1 |_{\Gamma_c} =0, \nu\times\textbf{V}_1 |_\Gamma =\nu\times \textbf{E}_2 |_\Gamma\ee
gives$$\nu\times\mbox{curl }\textbf{V}_1 |_\Gamma =
\nu\times\mbox{curl }\textbf{E}_2 |_\Gamma. $$
 Then by Green's
theorem again \be \label{a2} \ds 0 =\int_{\Omega} \textbf{E}_1\cdot (n_1 -n_1)\textbf{V}_1 \; dx
=\int_\Gamma [(\nu\times\mbox{curl }\textbf{E}_1) \cdot\textbf{V}_1 \ee
$$-\textbf{E}_1\cdot(\nu\times\mbox{curl }\textbf{V}_1)  ]\; ds  $$
$$
=-\int_\Gamma [(\nu\times\mbox{curl }\textbf{E}_1) \cdot(\nu\times\textbf{V}_1)
-(\nu\times\mbox{curl }\textbf{V}_1) \cdot(\nu\times\textbf{E}_1) ]\; ds.$$
The last relation may be deduced by a triple product (e.g., by using the Levi-Civita symbol we write $(\mathbf {b} \times \mathbf {c} )\cdot \mathbf {a}
=\varepsilon _{ijs}a^{i}b^{j}c^{s}$).\\ Then
from (\ref{eq1}), (\ref{a1}), and (\ref{a2}) we deduce the desired identity (\ref{i2}). \square\\

The second Lemma states that the set of solutions of the
Maxwell's equations with boundary data $0$ on $\Gamma_c$ is
dense in $\textbf{L}^2(\mathcal{O})$ in the set of all solutions.
\begin{lem}\label{lem2} Let $\it{\textbf{n}} \in L^\infty(\Omega)$. Let $\mathcal{O}$ be as in Lemma \ref{lem1} such that
$\Omega\backslash\overline{\mathcal{O}}$ is connected. Let us define $$\widetilde{S}(\Omega) = \Bigr\{\textbf{V}
\in \textbf{H}^2(\Omega) \, |\, (\frac{1}{k^2}\mbox{curl }\mbox{curl }- \it{\textbf{n}})\textbf{V}= 0 \ \mbox{in} \ \Omega, \nu\times\textbf{V}=
0 \ \mbox{on} \ \ \Gamma_c \Bigr\}$$ and $$S(\Omega) =  \Bigr\{\textbf{V}
\in \textbf{H}^2(\Omega) \, |\, (\frac{1}{k^2}\mbox{curl }\mbox{curl }- \it{\textbf{n}})\textbf{V}= 0 \ \mbox{in} \ \Omega\Bigr\}.$$ Then $\widetilde{S}(\Omega)$ is dense in $S(\Omega)$ according to $\textbf{L}^2(\mathcal{O})$ norm.
\end{lem}
\proof
We first define Green's function $\mathbf{G}(x, y)$ for (\ref{eq-maxwell1}) as a $3\times 3$ matrix valued
function solution of:\be\label{green-matrix}
  \left\{
\begin{array}{l}
(\mbox{curl }\mbox{curl }- k^2\emph{\textbf{n}}) \mathbf{G}(x,y) = -\mathbf{I}_3\delta(x-y) \ \ \mbox{in} \ \ \Omega,\\
 \nu\times \mathbf{G}(x,y) = 0 \ \ \mbox{on} \ \ x \in \partial \Omega\\
%  \lim_{|x|\to \infty}|x|\big[ \nabla\times G(x,x^{\prime})-ik\frac{x}{|x|}\times G(x,x^{\prime}) \big]=0
\end{array}
\right.
 \ee
where $\mathbf{ I}_3$ is the $3\times 3$ identity matrix. In
the above notation the curl operator acts on
matrices column by column. The Green function $\mathbf{G}$ is
given by
$$
\mathbf{G}(x,y)=\Phi(x,y)\big[
\mathbf{I}_3+ \ds\frac{\nabla_x\nabla_x}{k^2}  \big],
$$
where the scalar function $\Phi$ means the outgoing fundamental
solution for the Helmholtz operator "$\Delta +k^2$" and given by
$$
\Phi(x,y):=\ds\frac{e^{ik|x-y|}}{4\pi|x-y|}.
$$
As an example, the first column of $\mathbf{G}(x,y)$
equals

$$
\left(\begin{array}{c}
 \Phi(x,y) \\
  0 \\
  0
\end{array}
       \right)+\ds\frac{1}{k^2}\nabla_x[\nabla_x\cdot \left(\begin{array}{c}
\Phi(x,y) \\
  0 \\
  0
\end{array}\right).
$$

Multiplying equation~(\ref{eq-maxwell1}) by $\mathbf{G}(x,y)\cdot {\bf V }$ $({\bf V }\in \R^3)$,
integrating by parts in the domain $\Omega$, and using the relation (\ref{relation-E-H}) we immediately get a convenable integral
representation formula for the electric field called \emph{Stratton–Chu formula}. For more detail about this representation, one can see Theorem 6.1 in \cite{kress}.\\

Subsequently, suppose there exists $\textbf{V}\in S(\Omega)$ such that
 \be\label{eq-density1}\int_{\mathcal{O}} \textbf{V}\cdot\overline{\textbf{V}^{\prime} }dx = 0,\quad  \mbox{for all } \; \textbf{V}^{\prime} \in
\widetilde{S}(\Omega).\ee

%Then, by referring to \cite{CK-book1} and to \cite{CP} (for a closely idea ), we can see that the vector valued function
Define the vector valued function\begin{equation}\label{def-u-1}\displaystyle \textbf{W}(x):=-\int_{\mathcal{O}}\mathbf{G}(x,y)\overline{\textbf{V} }(y)dy\in \textbf{H}^2(\Omega).\end{equation}
Then, by referring to (\ref{green-matrix}) we find that
$$(\mbox{curl }\mbox{curl }- k^2\emph{\textbf{n}})\textbf{W}(x)=\left\{
\begin{array}{l}
\overline{\textbf{V} }\ \ \mbox{in} \ \ \mathcal{O},\\
\nm  0 \ \ \mbox{in} \ \  \Omega\backslash\overline{\mathcal{O}}.
\end{array}
\right.
 $$
 %$\nu\times\textbf{W}(x)=0,$ for $x\in\partial\Omega$. To do this we may rewrite
Moreover,  by (\ref{def-u-1}) we may write:
\begin{equation}\label{def-u-2}\displaystyle \nu(x)\times\textbf{W}(x)=-\int_{\mathcal{O}} \nu(x)\times \mathbf{G}(x,y)\overline{\textbf{V} }(y)dy.\end{equation}
Since for any $x\in\partial\Omega$, $\nu(x)\times \mathbf{G}(x, y) = 0$ ($\forall y\in\mathcal{O} $), we have $\nu\times\textbf{W}|_{\partial\Omega}=0$.\\
On the other hand, for all $\textbf{V}^{\prime} \in\widetilde{S}(\Omega)$ integration by parts yields
\be\label{eq-density2}\textbf{V}^{\prime}(x)=\int_{\partial\Omega} \nu(y)\times\mbox{curl }\mathbf{G}(x,y)\textbf{V}^{\prime }(y)\;ds(y) \ee
$$=\int_{\Gamma} \nu(y)\times\mbox{curl }\mathbf{G}(x,y)\textbf{V}^{\prime }(y)\;ds(y) ,\quad  \mbox{for all } \; x \in\Omega.$$
Hence, by inserting identity (\ref{eq-density2}) into (\ref{eq-density1}) we immediately get
$$\int_{\mathcal{O}} \nu(x)\times\mbox{curl }_{x}\mathbf{G}(x,y)\textbf{V}(y)\;ds(y)=0,\quad  \mbox{for all } \; x \in\Gamma,$$
which means that\be\label{eq-density3}\nu\times\mbox{curl }\textbf{W}=0,\quad \mbox{in }\Gamma.\ee
Now, by the unique continuation principe, it follows that $\textbf{W}(x)=0$ for all $x\in \Omega\backslash\overline{\mathcal{O}}$, and $\nu\times\mbox{curl }\textbf{W}=\nu\times\textbf{W}=0,\quad \mbox{on }\partial\mathcal{O}$.\\
On the other hand, by Green's formula we get
$$\int_{\mathcal{O}}|\textbf{V}|^2dx =\int_{\mathcal{O}}\textbf{V} \cdot \overline{\textbf{V} }dx =\int_{\mathcal{O}}(\mbox{curl }\mbox{curl }
- k^2\emph{\textbf{n}})\textbf{W}\cdot \overline{\textbf{V} }dx $$
$$ =
\int_{\mathcal{O}}\textbf{W}\cdot(\mbox{curl }\mbox{curl }- k^2\emph{\textbf{n}})\overline{\textbf{V}}dx=0. $$
Hence,  $\textbf{V} \equiv 0$ in $\mathcal{O}$. To achieve the proof of our density result, we can apply again the unique continuation principe to $\textbf{V}$ to find that $\textbf{V} \equiv 0$ in $\Omega$.  \square\\
%which proves the density lemma. \square\\
%%%%%%%%%%%%%%%%%%%%%%%%

{\bf Proof of Theorem  \ref{th1}.}
%%%%%%%%%%
Thanks to Sylvester and Uhlmann \cite{SU}, we can construct complex geometric-optics solution (CGO) for the Maxwell's equations (\ref{eq-maxwell1}). More precisely, they constructed their CGO solution to Schr$\ddot{o}$dinger's equation by looking for a solution in the form $u(x) = e^{i\xi.x}(1 + R_\xi(x))$ where $\xi\in\R^3$ satisfying $\xi.\xi= 0$ and $R_\xi$ decays like $O(1/|\xi|)$.\\
After that, Sun and Uhlman \cite{Uhlmann1} and Colton and Pavairanta \cite{CP} proved that the CGO solution of the Maxwell's equations may be of the form:
\be\label{CGO1}\textbf{V} = e^{x.\xi}[\eta+\Psi(x,\xi)];\quad \xi,\eta\in\C^3,\ee
with $\Psi\in H_\delta^1(\R^3)$ and $\Psi=O(1)$ as $|z|\to+\infty$ ($z$ is a distinguished direction from $\xi$). Here, $ L^2_\delta(\R^3)$ denotes the the Hilbert
space $$ \ds
\textbf{L}^2_\delta(\R^3)= \{ \textbf{g} \in \textbf{L}^2_{loc}(\R^3);   \int_{\R^3} (1 +
|x|^2)^\delta |\textbf{f}(x)|^2\; dx \; < + \infty \},\quad \mbox{for }-1 < \delta <
0,$$ and $\textbf{H}_\delta^1(\R^3)$ denotes the corresponding Sobolev space. Moreover, $\xi$ and $\eta$ are complex constant vectors satisfying $\xi\cdot\xi:=k^2$, and $\xi\cdot\eta=0$.\\
To explain the distinguished direction, we may refer to \cite{Uhlmann1} to write:\be\label{rel-Xi}\xi=s\rho+i\frac{l}{2}+g(s)\omega_1,\quad \mbox{and }\eta=l-i\frac{|l|^2}{2s}\omega_1+\frac{g(s)}{s}l\quad\quad (i^2=-1),
\ee
 where $s>0$, $l\in\R^3$, $\rho=w_1+iw_2$ for $w_i\in S^2$ such that $w_1.l=w_2.l=w_1.w_2=0$, and $\ds g(s):=\frac{|l|^2+4k^2}{4s+2\sqrt{4s^2+|l|^2}+4k^2}.$\\

From previous results (e.g., \cite{Uhlmann1}) and from (\ref{CGO1}), we can construct CGO solution of (\ref{eq-maxwell1}) in $\R^3$ as follows.
\begin{prop}\label{prop-CGO}Let $\it{\textbf{n}}\in C^2(\overline{\Omega})$ be as in (\ref{index}). Extend $\it{\textbf{n}}=\it{\textbf{n}}_0$ in $\Omega_e=\R^3\backslash\overline{\Omega}$. Let $\xi$ and $\eta$ be as in (\ref{rel-Xi}), and let $-1 < \delta < 0$. Then
there exist $\kappa_1=\kappa_1(\Omega) > 0$ and $r > 0$ such that if $s > r$ and
$$\|\it{\textbf{n}}-\it{\textbf{n}}_0\|_{W^{2,\infty}(\Omega)}\leq \kappa_1,
$$
then there is a unique solution of (\ref{eq-maxwell1}) in $\R^3$ of the form
\be \label{eqv0} \textbf{V} = e^{x\cdot \xi}[\eta +
\Psi_{\it{\textbf{n}}}(x, \xi)],\ee
 for  $|\xi|$ sufficiently large, with $\Psi_{\it{\textbf{n}}}\in H_\delta^1(\R^3)$ and $\Psi_{\it{\textbf{n}}}=O(1)$ as $s\to+\infty$.\\
\end{prop}

Concerning the proof of Proposition \ref{prop-CGO}, one can follow the proof of Theorem 1.6 in \cite{Uhlmann1} by making the necessary changes that needed in our problem here.\\
Now, from Proposition \ref{prop-CGO}, we can remark the following.
\begin{rem}\label{rem-CGO}From (\ref{rel-Xi}) and Proposition \ref{prop-CGO}, the vector valued function $\Psi_{\it{\textbf{n}}}$ decays to zero in certain
distinguished directions as $\xi\to+\infty$. In particular $\Psi_{\it{\textbf{n}}}=O(1)$ as $s\to+\infty$, and this suffices for our purpose to prove our main theorem.
\end{rem}

To proceed with the proof, we define $$\tilde{\emph{\textbf{n}}}_j=\left\{
\begin{array}{l}
\emph{\textbf{n}}_j\ \ \mbox{in} \ \ \Omega,\\
\nm  \emph{\textbf{n}}_0 \ \ \mbox{in} \ \  \R^3\backslash\overline{\Omega}.
\end{array}
\right.
 $$
Then by Proposition \ref{prop-CGO} for $j=1,2$ and for $-1<\delta<0$, there exist $\kappa_1^{(j)}>0$ and $r_j>0$ such that if
\be\label{condition-n-s}s > \tilde{r}=\max(r_1,r_2),\quad \mbox{and } \|\tilde{\emph{\textbf{n}}}_j-\emph{\textbf{n}}_0\|_{W^{2,\infty}(\Omega)}\leq \tilde{\kappa}_1,
\ee
where $\tilde{\kappa}_1=\min(\kappa_1^{(1)},\kappa_1^{(2)})$ we can construct solutions of
 the problem $(\mbox{curl }\mbox{curl } - k^2\tilde{\emph{\textbf{n}}}_j)\textbf{V}_j =0$ in $\R^3$ of the form
\be \label{eqv} \textbf{V}_j = e^{x\cdot \xi_j}[\eta_j +
\Psi_{\tilde{\emph{\textbf{n}}}_j}(x, \xi_j)],\quad j=1, 2\ee
 with $\Psi_{\tilde{\emph{\textbf{n}}}_j}(\cdot, \xi_j)
 \in \textbf{L}^2_\delta(\R^3)$. Moreover, by Remark \ref{rem-CGO} $\Psi_{\tilde{\emph{\textbf{n}}}_j}(x, \xi_j)$ decays to zero in certain
distinguished directions as $\xi\to+\infty$. Precisely, $\Psi_{\tilde{\emph{\textbf{n}}}_j}(x, \xi_j)=O(1)$ as $s\to+\infty$.\\

 %it can be easily justified that
 %\be
 %\label{eqv2}
%\ds  ||\Psi_{\tilde{n}_i}(\cdot, \xi_i)||_{\textbf{L}^2_\delta(\R^3)} \leq
 %\frac{C}{|\xi_i|}.
 %\ee

 On the other hand, from (\ref{rel-Xi}) we can expand that $\xi=sw_1+g(s)w_1+i(sw_2+\frac{l}{2})$. Then, we may define
 \be
 \label{eqv2-1}
\ds  \xi_j =-(-1)^{j}[s+g(s)]w_1 + i[\frac{l}{2} -(-1)^{j}sw_2],\quad \mbox{for }j=1, 2\quad  (i^2=-1);
\ee
 where $s$, $g(s)$, $w_1$, $w_2$ and $l$ given as in (\ref{rel-Xi}). Consequently, we have $\xi_1+\xi_2=il$.\\

%%%%%%%%%%%%%%%%%%%%%%
To complete the proof, we  write down,
$$
 \ds  \int_{\Omega} \textbf{V}_1|_{\Omega}\cdot(\tilde{\emph{\textbf{n}}}_1 -\tilde{\emph{\textbf{n}}}_2) \textbf{V}_2|_{\Omega} \; dx= \int_{\Omega} \textbf{V}_1|_{\Omega}\cdot(\emph{\textbf{n}}_1 -\emph{\textbf{n}}_2) \textbf{V}_2|_{\Omega} \; dx .$$
 Having $supp(\emph{\textbf{n}}_1-\emph{\textbf{n}}_2)\subset \mathcal{O}$, we get
 $$\int_{\Omega} \textbf{V}_1|_{\Omega}\cdot(\emph{\textbf{n}}_1 -\emph{\textbf{n}}_2) \textbf{V}_2|_{\Omega} \; dx=\int_{\mathcal{O}} \textbf{V}_1|_{\Omega}\cdot(\emph{\textbf{n}}_1 -\emph{\textbf{n}}_2) \textbf{V}_2|_{\Omega} \; dx .$$
Since $\textbf{V}_i|_{\Omega}\in S(\Omega)$, we can apply Lemma \ref{lem2} to state that for $i=1,2$, $\textbf{V}_i|_{\Omega}$ can be approximated by elements of $\widetilde{S}(\Omega)$ in in $L^2(\mathcal{O})$ norm.\\
Therefore,
$$
 \ds   \int_{\mathcal{O}} \textbf{V}_1|_{\Omega}\cdot(\emph{\textbf{n}}_1 -\emph{\textbf{n}}_2) \textbf{V}_2|_{\Omega} \; dx $$
may be approximated by
$$
 \ds  \int_{\mathcal{O}} \tilde{\textbf{V}}_1\cdot(\emph{\textbf{n}}_1 -\emph{\textbf{n}}_2) \tilde{\textbf{V}}_2 \; dx ,$$ where $\tilde{\textbf{V}}_i \in \textbf{H}^1(\Omega)$ solution of $$
\label{eq0*}\begin{array}{l} (\mbox{curl }\mbox{curl }- k^2\emph{\textbf{n}}_i)\tilde{\textbf{V}}_i= 0  \quad  \mbox{in }\Omega \\
 \nu\times \tilde{\textbf{V}}_i|_{\Gamma_c} = 0, \quad i=1, 2 .
\end{array}
$$
But, by Lemma \ref{lem1} we have $$
 \ds  \int_{\mathcal{O}} \tilde{\textbf{V}}_1\cdot(\emph{\textbf{n}}_1 -\emph{\textbf{n}}_2) \tilde{\textbf{V}}_2 \; dx=0 .$$
Thus,
\be\label{eqv3}
 \ds  \int_{\Omega} \textbf{V}_1|_{\Omega}\cdot(\emph{\textbf{n}}_1 -\emph{\textbf{n}}_2) \textbf{V}_2|_{\Omega} \; dx =\int_{\mathcal{O}} \textbf{V}_1|_{\Omega}\cdot(\emph{\textbf{n}}_1 -\emph{\textbf{n}}_2) \textbf{V}_2|_{\Omega} \; dx=
0. \ee
Next, suppose that we have (\ref{condition-n-s}). Then taking into account Proposition \ref{prop-CGO}, substituting (\ref{eqv}) into (\ref{eqv3}), using
(\ref{eqv2-1}), considering Remark \ref{rem-CGO}, and letting  $s\rightarrow + \infty$. We conclude by the Fourier integral theorem that:
\[
\ds \widehat{(\emph{\textbf{n}}_1 - \emph{\textbf{\textbf{n}}}_2)} (-l) = 0,\quad \mbox{$\forall$ }l\in\R^3.
\]
The hats denoting the Fourier transforms of the corresponding functions. The theorem is now
proved. \square

%%%%%%%%%%%%%%%%%%%%%%%%%%%%%%%%%%%%%%%%%%%%%%%
%%%%%%%%%%%%%%%%%%%%%%%%%%%%%%%%%%%%%%%%%%%%%%%%%%
\section{Reconstruction of $\emph{\textbf{n}}$}
%\textbf{au lieu de faire 3. Reconstruction  faire 3. Asymptotic: l'idee est de developpe q en fonction q0 (q=q0+f(l) avec f(l) to 0 en utilisant developpement asymp de Psi).}

Let $\tilde{\emph{\textbf{n}}} \in C^2(\bar{\Omega})$ be a known function. Assume that
$\emph{\textbf{n}}=\tilde{\emph{\textbf{n}}}$ almost everywhere in a neighborhood of $\partial \Omega$.
Denote $\mathcal{O} \subset \subset \Omega$, $\mathcal{O}$
bounded open with $C^2$-boundary containing $supp(\emph{\textbf{n}}-\tilde{\emph{\textbf{n}}})$. In this section we derive a formula for calculating $\emph{\textbf{n}}$ from
the local impedance map $Z_\emph{\textbf{n}}:
\widetilde{TH}_{\mbox{div}}^{\frac{1}{2}}(\Gamma) \rightarrow
 TH_{\mbox{div}}^{\frac{1}{2}}(\partial\Omega)$.

Assume that $Z_\emph{\textbf{n}}$ is known, then for
any $\textbf{E}, \textbf{V} \in \textbf{H}_{\mbox{div}}^1(\Omega)$ satisfying
$$\begin{array}{l} (\mbox{curl }\mbox{curl }- k^2\emph{\textbf{n}})\textbf{E}= 0  \quad  \mbox{in }\Omega \\
(\mbox{curl }\mbox{curl }- k^2\tilde{\emph{\textbf{n}}})\textbf{V}= 0  \quad  \mbox{in }\Omega \\
 \nu\times \textbf{E}|_{\Gamma_c} =  \nu\times \textbf{V}|_{\Gamma_c} =0,
\end{array}
$$
we have
\be
\label{eqr1} \ds \int_{\Omega} \textbf{E}\cdot(\emph{\textbf{n}} -\tilde{\emph{\textbf{n}}}) \textbf{V} \; dx =
 \int_\Gamma  \nu\times\textbf{E}\cdot (Z_\emph{\textbf{n}} - Z_{\tilde{\emph{\textbf{n}}}})( \nu\times\textbf{V}) \; ds , \ee
 where $Z_{\tilde{\emph{\textbf{n}}}}$ denotes the local impedance map
 associated to the refractive index $\tilde{\emph{\textbf{n}}}$.

%%%%%%%%%%%%%%%%%%%%%%%%%%%%%%%%%%%%%%%%%%%%%%%%

Extend $\emph{\textbf{n}}$ and $\tilde{\emph{\textbf{n}}}$ by $\emph{\textbf{n}}_0$ in $\R^3$. Let $\xi \in \C^3\setminus
\{0\}$ with $\xi \cdot \xi =0$. Define $\textbf{E}_\xi$ to be the
solution of
 \be \begin{array}{l}
 \label{eqr3} (\mbox{curl }\mbox{curl }-k^2\emph{\textbf{n}}_0)\textbf{E}_\xi =0 \quad \mbox{ in }
\R^3 \setminus \overline{\Omega} , \\
 \nm
 (\mbox{curl }\mbox{curl }- k^2\emph{\textbf{n}})\textbf{E}_\xi = 0  \quad \mbox{ in } \Omega ,
 \end{array}
 \ee
subject to the radiation condition
 \be \label{eqr4} e^{-
x\cdot \xi} \textbf{E}_\xi -\eta \in\textbf{ L}^2_\delta = \{ \textbf{f}: \int_{\R^3} (1 +
|x|^2)^\delta |\textbf{f}(x)|^2\; dx \; < + \infty \},\quad \mbox{for } -1 < \delta <0,\quad \eta\in\R^3. \ee

According to Proposition \ref{prop-CGO} and to Proposition 2.11 in \cite{Uhlmann1}, one can easily expand:
\be\label{rel-0}\textbf{E}_\xi(x)=e^{x\cdot \xi} [\eta+(d_1+\tilde{d}_1)\rho+d_2l+D/s+R],
\ee
where the scalar functions $d_1$, $\tilde{d}_1$, $d_2$, and the vector functions $D$ and $R$ satisfy respectively:
  $$ \begin{array}{l}
d_1=d_1(x,\rho,\hat{l}) |l |;\quad \|d_1\|<C, \\
 \nm
\ds \tilde{d}_1=\tilde{d}_1(x,\rho,s,l);\quad \lim_{s\to\infty}\|\tilde{d}_1\|=0, \\
 \nm
d_2=\sqrt{\frac{\mu}{\mu_{\infty}}}-1=0, \\
 \nm
D=D_0(x,\rho,\hat{l}) + D_1(x,\rho,\hat{l})|l | + D_2(x,\rho,\hat{l})|l |^2,\quad \|D_i\|<C; i=0, 1,2,\\
\nm
 \ds R=R(x,\rho,s,l);\quad \lim_{s\to\infty}s\|R\|=0,\\
 \nm
 \end{array}
$$
where $\hat{l}=\frac{l}{|l|}$, $C$ is a positive constant independent of $\delta$ and $n$.
\begin{rem}\label{rem-1}To simplify our method, we shall set \be\label{rel-1}A+G=\eta+(d_1+\tilde{d}_1)\rho+d_2l+D/s+R,
\ee
where $A:=\eta+d_1\rho+d_2l$ satisfies a transport equation type, and the remainder $G$ satisfies $\ds  \lim_{s\to\infty}\|G\|_{\textbf{L}^2_\delta}=0$.
\end{rem}
Therefore, from (\ref{rel-0})-(\ref{rel-1}) we get:
\[\ds\nu\times\mbox{curl }\textbf{E}_\xi|_{ \Gamma} =-(\xi\times\textbf{E}_\xi)\times\nu|_{ \Gamma}+ \nu\times\mbox{curl }(A+G-\eta)e^{x\cdot \xi}|_{ \Gamma} \quad \forall \; x \in \Gamma,\]
and the Jacobi identity immediately gives
\[\ds\nu\times\mbox{curl }\textbf{E}_\xi|_{ \Gamma} =-(\nu\times\textbf{E}_\xi)\times\xi|_{ \Gamma}
-(\xi\times\nu)\times\textbf{E}_\xi|_{ \Gamma}
+ \nu\times\mbox{curl }(A+G-\eta)e^{x\cdot \xi}|_{ \Gamma} \quad \forall \; x \in \Gamma.\]
Since
\[
\ds \nu\times\mbox{curl }\textbf{E}_\xi|_\Gamma (x) =
Z_n(\textbf{E}_\xi|_\Gamma)
\]
we obtain that $\textbf{E}_\xi$ solves the following equation on the open surface $\Gamma$:
\be
\label{eqr5} Z_n(\nu\times\textbf{E}_\xi|_\Gamma)+ \mathcal{N}_\xi(\nu\times\textbf{E}_\xi|_\Gamma)(x) = (A+G)e^{x\cdot \xi}, \quad \forall \; x \in \Gamma, \ee
where $\mathcal{N}_\xi$ is the operator defined by $\mathcal{N}_\xi:\widetilde{TH}_{\mbox{div}}^{\frac{1}{2}}(\Gamma)\to TH_{\mbox{div}}^{\frac{1}{2}}(\partial\Omega),\nu\times\textbf{f}\mapsto (\nu\times \textbf{f})\times\xi+C_\xi(\nu\times\textbf{f})$ with
 $C_\xi(\nu\times\textbf{f})=(\xi\times\nu)\times\textbf{f}$ is a bounded map on $\widetilde{TH}_{\mbox{div}}^{\frac{1}{2}}(\Gamma)$.\\
Then, the following holds.
\begin{prop} Assume that $k^2$ is not an eigenfrequency of $(\mbox{curl }\mbox{curl }- k^2\it{\textbf{n}})$ in $\Omega$.
Suppose that $\textbf{E}_\xi$ is a solution of (\ref{eqr3})-(\ref{eqr4}), then $\nu\times\textbf{E}_\xi |_\Gamma$ solves (\ref{eqr5}) uniquely.
\end{prop}

Now, let $\xi \in \C^3\setminus \{0\}$ with $\xi \cdot \xi =0$. Let
$\nu\times\textbf{E}_\xi |_\Gamma \in \widetilde{TH}_{\mbox{div}}^{\frac{1}{2}}(\Gamma)$ be the
solution of (\ref{eqr5}). Then, according to Section 2, we may have the following representation:
\be\label{rel-2}
\ds
 -\int_{\Omega} \nu\times(G(x,y)\ \textbf{E}_\xi) \ dy =      \nu\times e^{x\cdot \xi}(\eta +
\Psi_{\emph{\textbf{n}}}(x, \xi)),
\ee
where
\be\label{rel-2-0}
\ds \Psi_\emph{\textbf{n}}(x, \xi)=O(1)\quad \mbox{as }s\to+\infty.
\ee
Moreover, a carefully analysis on properties of operators
$Z_\emph{\textbf{n}}$ and  $\mathcal{N}_\xi$ immediately gives, by relation (\ref{eqr5}): \be\label{rel-3}\ds \nu\times\textbf{E}_\xi|_\Gamma=(Z_\emph{\textbf{n}}+ \mathcal{N}_\xi)^{-1}((A+G)e^{x\cdot \xi}), \quad \forall \; x \in \Gamma.\ee

Hens, we have the following reconstruction formula.
\begin{thm} \label{th2}
Let $\tilde{\emph{\textbf{n}}} \in C^2(\bar{\Omega})$ be a given function.  Assume that
$k^2$ is not an eigenfrequency of $(\mbox{curl }\mbox{curl }- k^2\it{\textbf{n}})$ in $\Omega$, and
$\it{\textbf{n}}=\tilde{\it{\textbf{n}}}$ almost everywhere in a neighborhood of $\partial \Omega$.
Then
\[ \ds \widehat{(\it{\textbf{n}} -\tilde{\it{\textbf{n}}})}(-l) = \lim_{s\rightarrow + \infty}
 \int_\Gamma   (Z_{\it{\textbf{n}}}+ \mathcal{N}_{\xi_1})^{-1}((A+G)e^{x\cdot \xi_1}|_{\Gamma})\cdot(Z_{\it{\textbf{n}}}
 - Z_{\tilde{\it{\textbf{n}}}}) \big((Z_{\it{\textbf{n}}}+ \mathcal{N}_{\xi_2})^{-1}((A$$
$$+G)e^{x\cdot \xi_2}|_{\Gamma})\big) \; ds(x).
 \]
\end{thm}
\proof
An major step of the proof was given in the previous approaches. Now, from (\ref{eqv2-1}) we may write$$\xi_1+\xi_2=il\quad (\mbox{for } i^2=-1).$$ By applying relation (\ref{rel-2}), we can pose $$
 \textbf{E}= \textbf{E}_\xi = e^{x\cdot \xi_1}(\eta_1 + \Psi_{\emph{\textbf{n}}}(x, \xi_1)),$$
and
 $$
 \textbf{V}= \textbf{E}_\xi= e^{x\cdot \xi_2}(\eta_2 + \Psi_{\tilde{\emph{\textbf{n}}}}(x, \xi_2)),$$
 with $\eta_1\cdot\eta_2=1$, to obtain from (\ref{eqr1})  and (\ref{rel-3}) that$$\ds \int_{\Omega} e^{x\cdot (\xi_1+\xi_2)}\big[\eta_1\cdot\eta_2+\eta_2\cdot\Psi_{\emph{\textbf{n}}}(x, \xi_1)+\eta_1\cdot\Psi_{\tilde{\emph{\textbf{n}}}}(x, \xi_2)$$
 $$+\Psi_{\emph{\textbf{n}}}(x, \xi_1)\cdot\Psi_{\tilde{\emph{\textbf{n}}}}(x, \xi_2)\big](\emph{\textbf{n}} -\tilde{\emph{\textbf{n}}})(x) \; dx$$
\be\label{rel-4} =
 \int_\Gamma  (Z_\emph{\textbf{n}}+ \mathcal{N}_{\xi_1})^{-1}((A+G)e^{x\cdot \xi_1}|_{\Gamma})\cdot(Z_\emph{\textbf{n}}\ee
  $$- Z_{\tilde{\emph{\textbf{n}}}}) \big((Z_\emph{\textbf{n}}+ \mathcal{N}_{\xi_2})^{-1}((A+G)e^{x\cdot \xi_2}|_{\Gamma})\big)\; ds(x).
$$

%\be\label{rel-3}\ds \nu\times\textbf{E}_\xi|_\Gamma=(Z_n+ \mathcal{N}_\xi)^{-1}((A+G)e^{x\cdot \xi}), \quad \forall \; x \in \Gamma.\ee

Now, by using (\ref{eqv2-1}), we immediately get $|\xi_i|\leq s$, for $i=1,2$ if $s>\tilde{r}$ (where $\tilde{r}$ given by Proposition \ref{prop-CGO}).\\ Thus, by  (\ref{rel-2-0}), Remark 3, the left hand side of relation (\ref{rel-4}) may be written as:
$$\ds  \lim_{s\rightarrow +
\infty}\int_{\Omega} e^{ix\cdot l}\big[\eta_1\cdot\eta_2+\eta_2\cdot\Psi_{\emph{\textbf{n}}}(x, \xi_1)+\eta_1\cdot\Psi_{\tilde{\emph{\textbf{n}}}}(x, \xi_2)+\Psi_{\emph{\textbf{n}}}(x, \xi_1)\cdot\Psi_{\tilde{\emph{\textbf{n}}}}(x, \xi_2)\big](\emph{\textbf{n}} -\tilde{\emph{\textbf{n}}})(x) \; dx$$
$$=\lim_{s\rightarrow +
\infty}\int_{\Omega} e^{ix\cdot l}\big[1+o(\frac{1}{s})+o(\frac{1}{s})+o(\frac{1}{s^2})\big](\emph{\textbf{n}} -\tilde{\emph{\textbf{n}}})(x) \; dx$$
$$=\widehat{(\emph{\textbf{n}} -\tilde{\emph{\textbf{n}}})}(-l)
.$$
The theorem now follows by considering the limit of expression (\ref{rel-4}) as $s\rightarrow +
\infty$.
\square
%%%%%%%%%%%%%%%%%%%%%%%%%%%%%%%%%%%%%%%%%%%%%%%%%%%
\section{Application: reconstruction of the
locations of small volume fraction perturbations of the refractive index}
The aim of this section is to apply the reconstruction procedure described in
Section $3$ for identifying the locations of small volume fraction
perturbations of the refractive index. Assume that $\Omega \subset \R^3$ contains a finite number of inhomogeneities, each of the
form $z_j + \alpha B_j$, where $B_j \subset \R^3$ is a bounded,
smooth domain containing the origin. The total collection of
inhomogeneities is
 $ \ds {\cal B}_\alpha
 = \ds \cup_{j=1}^{m} (z_j  + \alpha B_j)$ with $$(z_i  + \alpha B_i)\cap(z_j  + \alpha B_j)=\emptyset, \mbox{ for }i\neq j.$$
  The points $z_j \in \Omega, j=1, \ldots, m,$ which determine the
  location of the inhomogeneities, are assumed to satisfy the
  following inequalities:
\be
\label{f1}
 | z_j  - z_l | \geq c_0 > 0, \forall \; j \neq l  \quad
\mbox{ and } \mbox{ dist} (z_j, \partial \Omega) \geq c
> 0, \forall \; j,
\ee where $c$ is a positive constant.  Assume that $\alpha
>0$, the common order of magnitude of the diameters of the
inhomogeneities, is sufficiently small, that these
 inhomogeneities are disjoint,  and that
their distance to $\R^3 \setminus \overline{\Omega}$ is larger
than $c$. Let $\Gamma \subset
\partial \Omega$ be a given open subset of $\partial \Omega$.
Let $\emph{\textbf{n}}(x) \in C^2(\bar{\Omega})$ denote the unperturbed refractive index. We
assume that $\emph{\textbf{n}}(x)$ is known on a neighborhood of the boundary
$\partial \Omega$.  Let $\emph{\textbf{n}}_j(x) \in C^2(\overline{z_j+\alpha
B_j})$ denote the refractive index of the j-th inhomogeneity, $z_j+\alpha
B_j$. Introduce the perturbed refractive index
\begin{equation}
\emph{\textbf{n}}_\alpha(x)=\left \{ \begin{array}{*{2}{l}}
 \emph{\textbf{n}}(x),\;\;& x \in \Omega \setminus \bar {\cal B}_\alpha,  \\
 \emph{\textbf{n}}_j(x),\;\;& x \in z_j+\alpha B_j, \;j=1 \ldots m.
\end{array}
\right . \label{murhodef}
\end{equation}
Let us introduce the (perturbed) Maxwell equations in the presence of the
inhomogeneities ${\cal B}_\alpha$
\begin{equation}\label{rel-5}\begin{array}{l} (\mbox{curl }\mbox{curl }- k^2\emph{\textbf{n}}_\alpha)\textbf{E}_\alpha= 0  \quad  \mbox{in }\Omega \\
 \nu\times \textbf{E}_\alpha= \textbf{f}\in \widetilde{TH}_{\mbox{div}}^{\frac{1}{2}}(\Gamma), \quad  \mbox{on }\partial\Omega
\end{array}\end{equation}
%%%%%%%%%%%%%%%%%%%%%%%%%%%%%%

and define the local impedance map associated to
$\emph{\textbf{n}}_\alpha$ by :$Z_{\emph{\textbf{n}}_\alpha}(\textbf{f}) =\mbox{curl }\textbf{E}_\alpha\times\nu|_{\Gamma}$ for all $\textbf{f} \in
\widetilde{TH}_{\mbox{div}}^{\frac{1}{2}}(\Gamma).$ Let $\textbf{E}$ denote the solution
to the Maxwell equations with the boundary
condition $\textbf{E}\times\nu=\textbf{f}$ on $\partial \Omega$ in the absence of any
inhomogeneities and $Z_{\emph{\textbf{n}}}$ be  the local impedance map associated to
$\emph{\textbf{n}}$.
\begin{hyp}\label{hyp1} Throughout this section we suppose that:
the constant $k^2 = w^2\epsilon_0\mu_0$ is such that the natural weak
formulation of the problem (\ref{rel-5}), in the absence of any
inhomogeneities, has a unique solution.
\end{hyp}
The goal in this section is to identify efficiently, by using Theorem \ref{th2},
 the locations $\{z_j\}_{j=1}^m$ of the small
inhomogeneities ${\cal B}_\alpha$ from the knowledge of the
difference between the local impedance maps $\ds Z_{\emph{\textbf{n}}_\alpha} -
Z_{\emph{\textbf{n}}}$ on $\Gamma$.

Let $\textbf{V}$ be any function  in $\tilde{S}(\Omega)$, where $\tilde{S}(\Omega)$ is given by Lemma 2. Then by referring to \cite{AVV}, the following asymptotic formula (we shall not detail the proof, but we refer to the reference so quoted for closely techniques concerning a magnetic field $\textbf{H}_\alpha$) can
be derived :

\begin{thm}\label{thm-3}
Suppose (\ref{f1}), (\ref{murhodef}) and Hypothesis \ref{hyp1} are satisfied. There exists $0 < \alpha_0$ such that, given an arbitrary $\textbf{f}\in \widetilde{TH}_{\mbox{div}}^{\frac{1}{2}}(\Gamma) $, and any $0 <\alpha<\alpha_0$, the boundary value problem (\ref{rel-5}) has a unique (weak) solution $\textbf{E}_\alpha$. The constant $\alpha_0$ depends on the domains $B_j$, $\Omega$, the constants and the number $m$, but is otherwise independent of the points $z_j$; $j=1,\cdots,m$. Let $\textbf{E}$
denote the unique (weak) solution to the boundary value problem (\ref{rel-5}), in the absence of any
inhomogeneities. Then, for $\textbf{V}\in\tilde{S}(\Omega)$ we have:

\be\label{rel-6}\ds
\int_{\Gamma} \Bigr( Z_{\it{\textbf{n}}_\alpha}(\textbf{E}_\alpha\times\nu)\cdot \textbf{V}- \textbf{E}_\alpha\cdot Z_{\it{\textbf{n}}}(\textbf{V}\times\nu)\Bigr)ds(x) =
\alpha^3\sum_{j=1}^m [\it{\textbf{n}}(z_j) \ee
$$- \it{\textbf{n}}_j(z_j)]( M^j\textbf{E}(z_j))\textbf{V}(z_j)+
o(\alpha^4),
$$where $M^j=(m_{p,q}^j)_{1\leq p,q\leq3}$  is a $3\times3$ positive, symmetric, definite matrix (called the (rescaled) polarization tensor of the inhomogeneity set $B_j$) and the
remainder $o(\alpha^4)$ is independent of the set of points
$\{z_j\}_{j=1}^m$.
\end{thm}
\proof  The existence and uniqueness of solution to problem (\ref{rel-5}) is completely fixed in \cite{AVV}, when the solution is a magnetic field $\textbf{H}_\alpha$. Concerning our work here, one can use the well-known relation (\ref{relation-E-H})
%\be\label{E-H}\textbf{H}=\frac{1}{i\mu\omega}\mbox{curl }\textbf{E}
%\ee
to justify also the existence and uniqueness (weakly) of solution to problem (\ref{rel-5}) for $\textbf{E}_\alpha$.\\

We focus our attention, now, to justify (\ref{rel-6}). Regarding Theorem 1 in \cite{AVV}, the authors developed an asymptotic formula concerning the perturbation, $(\textbf{H}_\alpha - \textbf{H}_0) \times\nu|_{\partial\Omega}$ , in
the (tangential) boundary magnetic field, caused by the presence of the inhomogeneities ($\alpha\to0$). Based on (\ref{relation-E-H}), we may write\be\label{E-H-1}\textbf{H}\times\nu=\frac{1}{i\mu\omega}\mbox{curl }\textbf{E}\times\nu,\quad \mbox{on } \partial\Omega.
\ee
As we said before that we don't give a detail to the proof of this theorem. But, we may insert relation (\ref{E-H-1}) into the formula provided by Theorem 1 in \cite{AVV} p.774, and use $\nu\times(\textbf{E}_\alpha\times\nu)$ as the projection of $\textbf{E}_\alpha$ onto the tangent plane  of $\partial\Omega$. Thus, by using a vector triple product and by assumption in this paper that the permeability $\mu=\mu_0$ (fixed), we can rescale the polarization tensor and we may simplify the formula in the reference by using the definitions of both $Z_{\emph{\textbf{n}}_\alpha}$ and $Z_{\emph{\textbf{n}}}$  to get precisely (\ref{rel-6}).
\square\\

In order to get simple equations for the unknown parameters, namely, for the points
$\{z_j\}_{j=1}^m$ and  the values $\{\emph{\textbf{n}}_j(z_j) \}_{j=1}^m$, we may make suitable choices for the test
functions $\textbf{V}$ in $\tilde{S}(\Omega)$. Similar
idea was used in the literature, and the associated numerical experiments have been
successfully conducted in the case of the (piecewise constant)
conductivity problem with boundary measurements on all
of $\partial \Omega$.\\
According to Section 3, we may define \be\label{rel-6-1}\Lambda(\xi_1,\xi_2)= \ds
 \int_\Gamma   (Z_{\emph{\textbf{n}}_\alpha}+ \mathcal{N}_{\xi_1})^{-1}((A+G)e^{x\cdot \xi_1}|_{\Gamma})\cdot(Z_{\emph{\textbf{n}}_\alpha} - Z_{\emph{\textbf{n}}}) \big((Z_\emph{\textbf{n}} \ee
$$+ \mathcal{N}_{\xi_2})^{-1}((A+G)e^{x\cdot \xi_2}|_{\Gamma})\big) \; ds(x).
$$

From Theorem \ref{th2}
 and Sobolev's embedding
theorem we can take
\[
\begin{array}{l}
\ds  \textbf{E}(z_j) = e^{z_j\cdot \xi_1}(\eta_1 + \Psi_{\emph{\textbf{n}}}(x, \xi_1)) \\
\nm \ds \textbf{V}(z_j) = e^{z_j\cdot \xi_2}(\eta_2 + \Psi_{\emph{\textbf{n}}}(x, \xi_2))
\end{array}
\]
with $\eta_1\cdot\eta_2=1$ and $\xi_1+\xi_2=il$ (for $i^2=-1$) to obtain from (\ref{rel-6}) that
%\newpage
\be\label{rel-7}\ds \Lambda(\xi_1,\xi_2) =\ds
\int_{\Gamma} \Bigr( Z_{\emph{\textbf{n}}_\alpha}(\textbf{E}\times\nu)\cdot \textbf{V}- \textbf{E}\cdot Z_{\emph{\textbf{n}}}(\textbf{V}\times\nu)\Bigr)ds =\alpha^3\sum_{j=1}^m (\emph{\textbf{n}}(z_j)
\ee
$$- \emph{\textbf{n}}_j(z_j)) M^je^{il\cdot z_j}
+
o(\alpha^4).
$$
Then, by neglecting the remainders $ o(\alpha^4)$ in (\ref{rel-7}) we may achieve the proof of the following result.
\begin{cor}\label{cor-1} Suppose that we have all hypothesis of Theorem 3. Let $\Lambda(\xi_1,\xi_2) $ be defined by (\ref{rel-6-1}). Then, the locations $\{z_j\}_{j=1}^m$ are obtained as supports of the inverse Fourier
transform of $\Lambda(\xi_1,\xi_2) $.
\end{cor}

Finally, it follows from Corollary \ref{cor-1} that the centers $\{z_j;\quad j=1,\cdots,m\}$ can be recovered easily, and therefore the values $\emph{\textbf{n}}_j(z_j)$ (for $j=1,\cdots,m$) could be obtained by solving a linear system arising
from (\ref{rel-7}). The extension to general geometries, and /or to anisotropic domain, would allow us to
deal with real-life applications. This may be considered in further paper.

%\section*{Acknowledgements}

\end{document}